\documentclass[preprint,preprintnumbers,amsmath,amssymb]{revtex4}
\usepackage{amsfonts}    
\usepackage{amssymb}
\usepackage{latexsym}
\usepackage{graphicx}
\usepackage[usenames,dvipsnames]{color}

\newcommand{\al}{\alpha}

\newcommand{\la}{\lambda}

\newcommand{\De}{\Delta}

\newcommand{\rar}{\rightarrow}

\begin{document}

\title{Nuclear critical charge for two-electron ion in Lagrange mesh method}

\author{H.~Olivares Pil\'on}
\email{horop@nucleares.unam.mx}
\affiliation{Departamento de F\'isica, Universidad Aut\'onoma Metropolitana-Iztapalapa, Apartado Postal 55-534, 09340 M\'exico, D.F., Mexico}

\author{A.V.~Turbiner}
\email{turbiner@nucleares.unam.mx}
 \affiliation{Instituto de Ciencias Nucleares, Universidad Nacional
 Aut\'onoma de M\'exico, Apartado Postal 70-543, 04510 M\'exico,
 D.F., Mexico}

\begin{abstract}
The Schroedinger equation for two electrons in the field of a charged fixed center $Z$ is solved with the Lagrange mesh method for charges close to the critical charge $Z_{cr}$. We confirm the value of the nuclear critical charge $Z_{cr}$ recently calculated in Estienne et al. {\em Phys. Rev. Lett. \bf 112}, 173001 (2014) to 11 decimal digits using an inhomogeneous (non-uniform) three-dimensional lattice of size $70 \times 70 \times 20$. We show that the ground state energy for H$^-$ is accurate to 14 decimals on the lattice $50 \times 50 \times 40$ in comparison with the highly accurate result by Nakashima-Nakatsuji, {\it J. Chem. Phys. \bf 127}, 224104 (2007).

\date{November 8, 2014}

\vskip 2cm


\end{abstract}

\pacs{31.15.Pf,31.10.+z,32.60.+i,97.10.Ld}

\maketitle

The three-body Coulomb problem is a fundamental problem of quantum mechanics. A number
of three-body Coulomb systems exist in Nature: two-electron atomic ions and one-electron hydrogen-like molecular ions are among two of them. The simplest version of (reduced) three-body Coulomb system is the so called helium-like sequence $(Z, e, e)$, where the charge $Z$ is assumed infinitely-heavy. There are two important characteristics of the system: (i) the {\it critical} charge $Z_{cr}$ for which the ionization energy vanishes and the ground state level touches (hits, kicks) continuum or, in other words, the ground state level degenerates with the threshold of continuum and (ii) the so called {\it bound} charge $Z_B$ such that at $Z < Z_B$ the Schroedinger equation has no normalizable solutions and, likely, the ground state energy $E(Z)$ has a singularity at $Z = Z_B$ (for discussion see \cite{Stillinger:1966}). Following W.~Reinhardt conjecture based on a concept of delatation analyticity of Coulomb Hamiltonians \cite{Reinhardt:1977}, these charges have to be equal $Z_{cr}=Z_B$. Recently, Estienne et al. \cite{Drake:2014} in highly precise variational calculation obtained the accurate value for the critical charge
\begin{equation}
\label{zcrit}
      Z_{cr}\ =\ 0.911\, 028\, 224\, 077\, 255\, 73\ ,
\end{equation}
{\it significantly} improving all previous results, thus ending a controversy which existed in literature. The aim of this Letter is to verify this result by making an alternative calculation.

Before to proceeding it is important to discuss physics related the critical charge $Z_{cr}$.
It is evident from the theoretical point of view, that the critical charge can exist for a three body Coulomb system $(Z, e, e)$ in full dynamics, for finite mass $m_Z$ of charge $Z$. This mass depends strongly on $Z$. In Nature the charge $Z$ takes integer values $Z=1,2,3,\ldots$ as well as its mass $m_Z$ takes some discrete values $m_p, m_{\al}, m_{li} \ldots $ which are always very large compared to the electron mass. Since it is quite unlikely that the system $(Z_{cr}, e, e)$ does exist in Nature, we do not know what a value of mass $m_{Z_{cr}}$ has to be assigned to the critical charge, except for it being large, say, of the order of proton mass.
Conceptually, it gives no chance to find finite-mass corrections to $Z_{cr}$.
All this justifies a consideration of the system $(Z, e, e)$ in the static limit, $m_Z =\infty$.
However,  even in static approximation there exist various corrections to $Z_{cr}$ due to the relativistic effects, the effects of non-QED interactions etc. There are also corrections due to limited (to 8-9-10 s.d.) experimental knowledge of atomic units at present. Thus, from physical point of view it does not make much sense to calculate $Z_{cr}$ with very high accuracy. This is why we decided to limit our accuracy in determination of $Z_{cr}$ to 11 s.d.

The Lagrange-mesh method ~\cite{BH1986, HB2003} is a numerical procedure for solving the Schroedinger equation by placing it into a non-uniform inhomogeneous lattice defined by zeroes of classical orthogonal polynomials using a basis of Lagrange functions and the associated Gauss quadratures. So far in all applications this method demonstrated very fast convergence: lattice size of a few dozens was sufficient to get 11-12 significant digits (s.d.) using double precision arithmetic with very modest computational efforts.
For instance, 12 s.d. have been obtained in the rotational-vibrational energies of the ground state for the three body system  H$_2^+$ \cite{OB12H2} (see also \cite{TO:2011}) and some of their isotopomers: D$_2^+$~\cite{OP13D2}, HD$^+$~\cite{OB13HD} and T$_2^+$~\cite{OP14T2} in full dynamics (beyond Born-Oppenheimer approximation). In this approach, in the center-of-mass coordinate system, the wave function is presented as a sum of products of the Wigner matrix elements with coefficient functions depending only on the relative coordinates of the three particles. Finding these functions leads to a system of coupled differential equations which are placed on a (non-uniform) lattice. Thus, the relative motion is described as a product of three one-dimensional Lagrange-Laguerre functions each one of them being an expansion containing $N_x$, $N_y$ and $N_z$ terms, respectively (see \cite{HB2003,OB12H2} for details). In particular, in \cite{HB2003,OB12H2} it was shown that the procedure converges with the increase of number of mesh points $N_x$, $N_y$ and $N_z$. Furthermore, the rate of convergence in mesh sizes $N_x$, $N_y$ and $N_z$
can be changed if three scaling parameters $h_x$, $h_y$ and $h_z$ are introduced in directions $x,y,z$ , respectively. These parameters control the size of domain in configuration space where the lattice is introduced adjusting effectively the basis to the size of the system. It can be checked that convergence is reached at any reasonable value of parameters in $h \in [0.1 - 3.]$.
However, usually, there exists a domain of variation of values of these parameters leading to  faster convergence. In practice, we are interested in finding the minimal size of the mesh which provides the desired number of significant digits in energy.
In this Letter we will study the two-electron system $(Z, e, e)$ in the static approximation assuming the charge $Z$ is infinitely massive and its value is close to critical using the Lagrange-mesh method in the form presented in \cite{OB12H2}.

Two electron system with infinitely-massive charge center $Z$ is described by the Hamiltonian
\begin{equation}
\label{H}
    {\cal H}\ =\ -\frac{1}{2} (\De_1 + \De_2) \ -\ \frac{Z}{r_{1}}\ -\
    \frac{Z}{r_{2}}  \ +\ \frac{1}{r_{12}}\ ,
\end{equation}
written in atomic units with electron mass $m_e=1$.
A change of variables in (\ref{H}), $\vec{r} \rar \vec{r}/Z$,\ leads to a new form of the Hamiltonian
\begin{equation}
\label{H_t}
    {\cal H}_t\ =\ -\frac{1}{2} (\De_1 + \De_2)\ -\ \frac{1}{r_{1}}\ -\
    \frac{1}{r_{2}}  \ +\  \frac{\la}{r_{12}}\ , \quad \la = \frac{1}{Z}\ ,
\end{equation}
where the new energy $\tilde E(\la)=\frac{E(Z)}{Z^2}$, while $E(Z)$ is the ground state energy of (\ref{H}). The Lagrange-mesh method is applied to the Schroedinger equation for the Hamiltonian (\ref{H_t}),
\[
     {\cal H}_t\ \Psi \ =\ \tilde E(\la) \Psi \ .
\]

The numerical calculations are carried out using code written in FORTRAN in double-precision arithmetics. The code includes the program JADAMILU designed for fast diagonalization of large sparse matrices \cite{Boll:2007}. The calculations were performed in cluster KAREN using a single 2.8 GHz core (ICN-UNAM).
Convergence was checked with mesh size increase in each direction until reaching stability of the desired number of digits in energy. This procedure was repeated for different values of scale parameters $h_x$, $h_y$ and $h_z$. Following computational convenience we limit the size of lattice by $70 \times 70 \times 20$ in $x,y,z$ directions respectively.
Eventually, for given $Z$ some suitable values of $h_x$, $h_y$ and $h_z$ were found which lead to maximal number of significant digits (s.d.) in energy for the lattice $70 \times 70 \times 20$. This number of s.d. varied from 14 for $Z=1$ to 11 for $Z=Z_{cr}$.
It is clear that such suitable values of $h_x$, $h_y$ and $h_z$ depend on $Z$ since the domain of configuration space giving dominant contribution changes with $Z$. This dependence can be very strong especially when we approach the critical charge: the electronic distribution changes dramatically at the critical charge (for discussion see e.g. \cite{Simon:1977}).

For a given charge $Z$, the process of checking convergence and finding suitable values for scale parameters took several hours of CPU time while the final run at maximal mesh sizes $70 \times 70 \times 20$ took 20-30 minutes of CPU time.

\begin{table}
\begin{tabular}{|l| l | l | l |}
\hline\hline
$Z$          &\quad       $E$ (a.u.) \ &\  present\ &\ $\quad [N_x=N_y, N_z; h_x=h_y, h_z]$ \\
\tableline
\ 1.00  \  &\quad  -0.527 751 016 544 377$^{a}$ \ &\quad  -0.527 751 016 544 38 \ &\quad [50, 40;
0.8, 0.5]\\
\ 1.00$^{\star}$ \ &\quad -0.527 445 881 114 178$^{aa}$ \ &\quad -0.527 445 881 114 18 \ &\quad [50, 40;
0.8, 0.5]\\
\ 0.95  \  &\quad  -0.462 124 684 390 $^{b}$ \ &\quad  -0.462 124 699 683 8 \ &\quad [50, 40 ;1.0, 0.5]\\
\ 0.92  \  &\quad  --  \ &\quad  -0.425 485 281 676  \   &\quad [70, 20; 1.0, 0.6]\\
\ 0.911\, 028\, 224\, 07(7)  \ &\quad   \ &\quad  -0.414 986 212 53 \ &\quad [70, 20; 2.4, 0.4]\\
\ $Z^{EBMD}_{cr}$ (see (\ref{zcrit})) \ &\quad  -0.414 986 212 532 679 $^{c}$ \ &\quad  -0.414 986 212 53 \ &\quad [70, 20; 2.4, 0.4]\\
\tableline
\hline
\end{tabular}
\caption{
\label{table1}
 Ground state energy $E$ for a two-electron system for selected values of $Z$ compared
 with \cite{Nakashima:2007}$^{a}$, \cite{TG:2011}$^{b}$, \cite{Drake:2014}$^{c}$ (rounded, see text) for infinite massive charge $Z$. At $Z=1$ and for proton mass, $m_p=1836.152701 m_e$ marked by ${\star}$ the energy is calculated and compared
 with \cite{Ackermann:1996}$^{aa}$. Lattice size and scale parameters are indicated.}
\end{table}

The results of calculations are presented in Table \ref{table1}.
First of all, let us note that ground state energy of the negative hydrogen ion, $Z=1$ \footnote{Also presented in~\cite{HB1999} using the Lagrange-mesh approach but with questionable 14th decimal after rounding} coincides in 14 decimal digits with highly accurate result obtained in \cite{Nakashima:2007} in the static limit, $m_p=\infty$, as well as for the case of finite proton mass in \cite{Ackermann:1996}.

Estienne et al. \cite{Drake:2014} (EBMD) using a variational calculation with triple basis sets containing up to 2276 terms obtained the value for the critical charge
\begin{equation}
\label{zcrit}
      Z^{EBMD}_{cr}\ =\ 0.911\, 028\, 224\, 077\, 255\, 73\ ,
\end{equation}
which corresponds to threshold energy (the lower level of continuum),
\begin{equation}
\label{th}
E_{th}^{EBMD} = - \frac{(Z^{EBMD}_{cr})^2}{2}\ .
\end{equation}
In order to find the critical charge in the present approach we have to calculate the ionization energy, $I(Z)= E(Z) + Z^2/2$, {\it vs} $Z$, and look for a charge for which it vanishes, $I(Z_{cr})=0$.
Solving the Schroedinger equation using the Lagrange-mesh method we were able to find the value of the charge $Z$ (see Table \ref{table1}) for which the ionization energy is of the order $10^{-12}$. It coincides with $Z^{EBMD}_{cr}$, see (\ref{zcrit}), and their threshold energy $E_{th}^{EBMD}$, see (\ref{th}), in 11 s.d. after rounding. Independently, we calculated the threshold energy $E_{th}$ at $Z=Z^{EBMD}_{cr}$ and compared with $E_{th}^{EBMD}$: it also shows coincidence in 11 s.d. All this implies the coincidence in the value of critical charge with one found in \cite{Drake:2014} in 11 s.d. (after rounding) \footnote{It was checked that the increase of the mesh up to $85 \times 85 \times 20$ allows us to reproduce at least 12 decimal digits in (1), (4) using the same values of parameters $h_x$, $h_y$ and $h_z$ as in Table \ref{table1}.}.
The result for $Z=0.95$ obtained in \cite{TG:2011} using Korobov functions method \footnote{The size of the basis was chosen in such a way to reproduce 12 s.d. in the ground state energy of H$^-$ (see for discussion \cite{TG:2011})} for the ground state energy was also checked: it starts to differ from the present calculation in the 8th decimal digit(!). It indicates a slower convergence of the Korobov method at $Z < 1$ than naively expected (see a discussion in \cite{TG:2011}, \cite{Drake:2014}).

To summarize, with an alternative calculation we confirm the value of the critical nuclear charge $Z_{cr}$ obtained in \cite{Drake:2014} in 11 significant digits. Preliminary calculations of the ground state energy {\it vs.} $Z$ in a close vicinity to $Z_{cr}$ performed with accuracy of 11 significant digits, do not give an indication to the presence of a singularity. A detailed study will be performed elsewhere.

\textit{\small Acknowledgements}.  HOP wants to express his gratitude to D Turbiner (Stanford) for help with computer calculations.  HOP also thanks the department PNTPM (Universite Libre de Bruxelles) for the use of their computer facility where preliminary calculations were carried out.

The research by AVT is supported in part by DGAPA grant IN109512 (Mexico). AVT thanks the University Program FENOMEC (UNAM, Mexico) for partial support.

\end{document}